\documentclass[a4paper,11pt]{article}
\pdfoutput=1 

\usepackage{jinstpub} 

\usepackage{amsmath}
\usepackage{amsfonts}
\usepackage{amssymb}

\usepackage{lineno}

\title{A comparison between scintillation light Analog and Digital trigger 
for large volume Liquid Argon Time Projection Chambers}

\author[a,b]{O.~Barnab\`a,}
\author[a,b]{A.~Menegolli,} 
\author[a,b]{R.~Nard\`o,}
\author[a,b]{M.~Pirola,}
\author[a]{M.C.~Prata,} 
\author[a,1]{G.L.~Raselli\note{Corresponding author.},}
\author[a,b]{E.~Romano}
\author[a]{and M.~Rossella}

\affiliation[a]{INFN, Pavia, Italy}
\affiliation[b]{University of Pavia, Pavia, Italy}

\emailAdd{gianluca.raselli@pv.infn.it}

\abstract{
Large volume Liquid Argon Time Projection Chambers (LAr-TPC) are used and proposed for neutrino physics and rare event search. Most of these detectors make use of the scintillation light of liquid argon for trigger purposes. 
Two different approaches can be adopted to provide these detectors with an effective trigger system, relying upon analog or digital processing of signal coming from photodetectors, like photomultiplier tubes or silicon photomultipliers. Each method presents advantages and drawbacks, so the implementation of a hybrid solution 
can benefit from both approaches. 
To this purpose, an innovative electronic board prototype has been designed and proposed for the use in large volume LAr-TPC detectors.}


\proceeding{15$^{\text{th}}$ Topical Seminar on Innovative Particle and Radiation Detectors (IPRD19) \\
  14-17 October 2019\\
  Siena, Italy}

\begin{document}

\maketitle
\flushbottom
\section{Introduction}

Large volume Liquid Argon Time Projection Chambers (LAr-TPC) are extensively used and proposed for neutrino physics and rare event search. 
Most of these detectors exploit the scintillation light of liquid argon for trigger purposes. 
Light is usually detected by photodetectors, like photomultiplier tubes (PMT) or silicon photomultipliers (SiPM), which provide
fast electrical pulses whose shape is directly related to the number and time distribution of the collected scintillation photons.
Two different approaches are adopted to provide LAr-TPC detectors with an effective trigger system based on the scintillation light.
The first one, in this paper named {\em Analog Trigger Approach}, is based on the analog processing of signals coming from individual photodetectors
in defined sectors of the experimental apparatus,
followed by a digitization/discrimination to produce trigger gates.
In the second one, named {\em Digital Trigger Approach}, signals coming from each individual photodetector are digitized/discriminated by fast
electronics and logically combined by logic units to generate trigger gates.

Purpose of this paper is to describe and compare these approaches together with a presentation of the implementation of
an innovative electronic board which allows the hybrid exploitation of these two techniques for the use in large
volume LAr-TPC detectors.
The main characteristics of the  LAr scintillation light are briefly summarized in section~\ref{sec:light}.
Section~\ref{sec:trigger} describes the two approaches presently adopted to exploit the scintillation light
for trigger purposes in LAr-TPC. Two actual examples are presented in section~\ref{sec:example}. 
Finally, the implementation of a custom designed board for the hybrid exploitation of the two approaches is presented in section~\ref{sec:board}.

\section{Scintillation light}
\label{sec:light}

The interaction of an ionizing particle in LAr produces a scintillation light in the vacuum
ultra violet region (VUV) emitting photons with a wavelength of 128~nm corresponding
to an energy of 9.7~eV (see \cite{CENNINI1999240} and references therein).
During the collision, the interacting particle excites and ionizes
the surrounding argon atoms, producing excimers Ar$^*_2$ (excited molecules) which decay
radiatively into VUV scintillation light. There are two processes forming Ar$^*_2$,
both ending up with the same radiative reaction.
Excitons Ar$^*$ (excited atoms) directly interact with the neighboring argon atoms,
forming excimers:

$$\rm  Ar^* + Ar \to Ar^*_2 \to 2Ar + \gamma \; (128~nm)$$
Ionized atoms can also form excimers via a recombination between Ar$^+_2$ and the
thermalized secondary electrons:

$$\rm  Ar^+ + Ar \to Ar^+_2 + e^- \to Ar^{**} + Ar  \to Ar^{**}_2 \to Ar^* + Ar + heat $$
$$\rm  Ar^* + Ar \to Ar^*_2 \to 2Ar + \gamma \; (128~nm)$$

Excimers are produced in two singlet states, $\rm ^1\Sigma ^-_u$ and $\rm ^1\Sigma ^+_u$, 
and in a triplet state $\rm ^3\Sigma ^-_u$.
The singlet state $\rm ^1\Sigma ^-_u$ does not emit photon because of the parity conservation. Therefore
the scintillation light possesses two components: those stemmed from the transitions
$\rm ^1\Sigma ^+_u \to  {^1\Sigma ^+_g}$ (fast decay) and $\rm ^3\Sigma ^+_u \to \rm {^1\Sigma ^+_g}$ (slow decay) 
where $\rm ^1\Sigma ^+_g$ is the ground state.
The decay of the singlet state is strongly allowed and its decay time $\tau_1$ is of the order of nanoseconds.
The triplet state has a longer lifetime because of the strong spin-orbit coupling in Ar$_2$ and
its decay time $\tau_2$  is measured to be $\approx 1.6$~$\mu$s. 
The populations of the singlet and triplet state depend on the ionization
density and, in turn, on the interacting particle type. Particle discrimination in
LAr is based on the relative intensity of the singlet and triplet states which results from
the ionization density.

\section{Trigger approaches}
\label{sec:trigger}

Two different approaches are presently adopted to exploit the LAr scintillation light for triggering
purposes, named in this paper: {\em Analog Trigger Approach} and {\em Digital Trigger Approach}.
The two approaches can be individually adopted, or can be combined in a hybrid configuration or even 
implemented at different
trigger levels. In the following the main advantages of the two techniques are described.

\subsection{Analog Trigger Approach}

\begin{figure}[h]
\centering
\includegraphics[width=0.75\columnwidth]{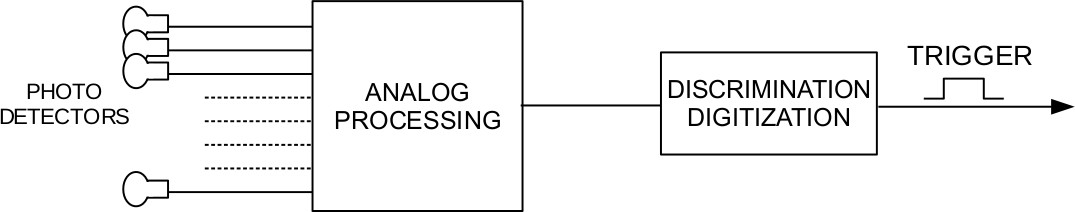}
\caption{{\em Analog Trigger Approach}: conceptual diagram.}
\label{fig:analog}
\end{figure}

This approach, sketched in figure~\ref{fig:analog}, is conceptually based on an analog processing of the signals coming from individual photodetectors.
Usually shaping preamplifiers with defined integration/differentiation parameters are used and the resulting signals are 
linearly summed up before the digitization which generates the trigger gates.
This technique is able to exploit both the fast and slow
components of the LAr scintillation light, making it suitable for the detection of low energy ionizing
events. An appropriate selection of integration/differentiation constants permits the recognition
of the nature of the interacting particle at trigger level by means of Pulse Shape Discrimination (PSD) techniques.
These features make this technique particularly suitable for LAr-TPC detectors for dark matter research,
where low energy nuclear recoils should be distinguished from
background generated by minimum ionizing particles. 

The main drawbacks of this approach are: $a)$ an interaction time information with coarse resolution due to the
analog sum of different photodetectors with different time response or because of the adoption of long integration constants
to exploit the delayed scintillation photons; $b)$ an increase of the electronic noise due to the linear sum
of photodetector outputs, especially when synchronous noise is present; $c)$ the need to develop low-noise
custom designed electronics to fit the apparatus requirements.

\subsection{Digital Trigger Approach}

\begin{figure}[h]
\centering
\includegraphics[width=0.75\columnwidth]{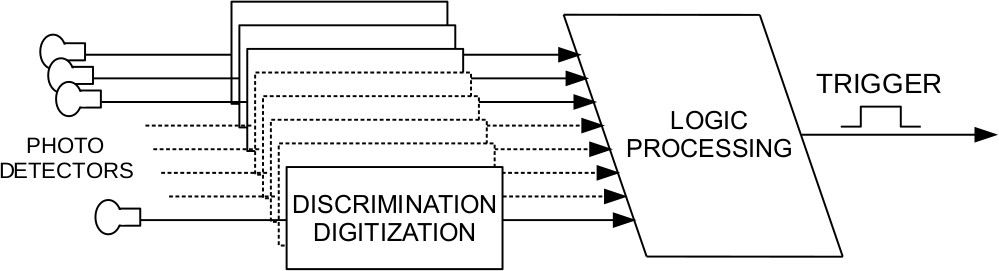}
\caption{{\em Digital Trigger Approach}: conceptual diagram.}
\label{fig:digital}
\end{figure}

This approach, sketched in figure~\ref{fig:digital}, is conceptually based on the digitization, by means of fast transient digitizers, of
signals coming from each individual photodetector in defined sectors of the experimental apparatus. 
The resulting digital pulses are then combined by programmable logic units to generate a trigger gate based on coincidence, 
majority and/or multiplicity of the available inputs. The development of custom designed electronics 
is usually not required and commercial standard electronic boards
can fit the apparatus requirements.
Only the fast component of the scintillation light is 
generally exploited in this technique
making possible to attain time resolution 
of the order of 100~ps.
On the contrary its implementation is only adequate
for the detection of ionizing interactions with high energy deposition 
and/or when the number of photodetectors is relatively high. 
This approach is useful 
for low-energy background rejection, such that produced by $^{39}$Ar or other radioactive
nuclides, composed of single sparse photons whose amplitude
does not individually exceed the discrimination threshold level.

\section{Example of different trigger systems}
\label{sec:example}

In the following two actual implementation instances of trigger system in LAr-TPC detectors are presented.
ICARUS T600 detector at Laboratori Nazionali del Gran Sasso (LNGS) was operated with a plain {\em Analog Trigger}\/ scheme, while the 
2.3~L WARP detector prototype adopted a hybrid solution with the two approaches at different trigger levels. 

\subsection{The ICARUS T600 detector at LNGS}

ICARUS T600 detector is made of two identical cryostats, filled with about 760~t of ultra-pure liquid argon~\cite{AMERIO2004329}. 
Each cryostat houses two TPCs with 1.5~m maximum drift path, sharing a common
central cathode made of punched stainless-steel panels. 

The detector took data from 2010 to 2012 at LNGS recording CERN Neutrinos to Gran Sasso (CNGS) beam and atmospheric neutrino interactions~\cite{Rubbia_2011}.
The ICARUS trigger system was based on the detection of LAr scintillation light by means of
74 ETL9357FLA PMTs with a diameter of 8~inches mounted
behind the wire chambers~\cite{ANKOWSKI2006146}. The sand-blasted
glass window of each device was coated with 200~$\mu$g/cm$^2$ of Tetraphenyl-Butadiene (TPB) to 
convert the VUV photons to visible light. 

\begin{figure}[h]
\centering
\includegraphics[width=0.95\columnwidth]{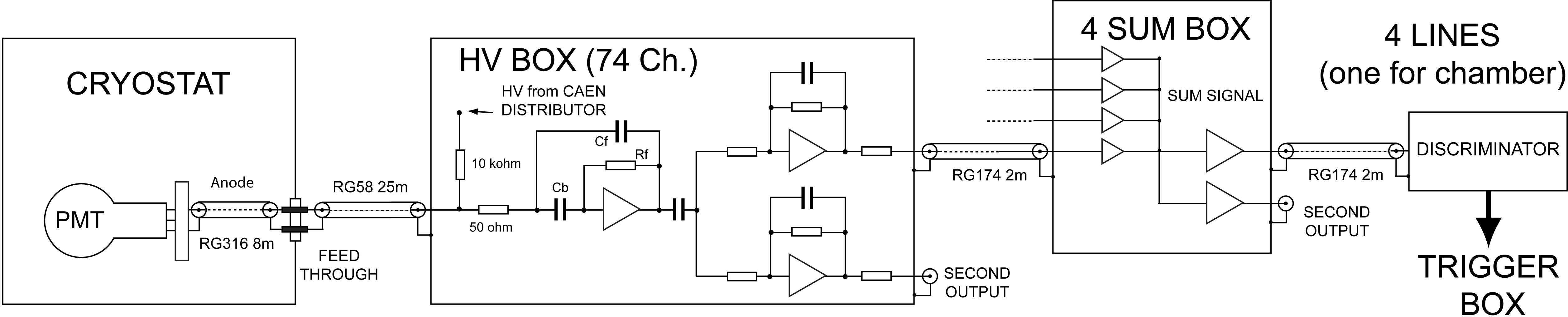}
\caption{Basic electronic scheme adopted for the ICARUS T600 detector at the LNGS.}
\label{fig:lngs}
\end{figure}

Since a low number of PMTs was installed, an analog trigger approach was adopted to exploit as much as possible
photons generated by ionizing interactions~\cite{Antonello_2014}.
The basic electronic scheme is shown in figure~\ref{fig:lngs}. 
For each channel a custom designed electronic circuit was realized to pick-up
and shape the anodic signal.  
The integration time constant was adjusted to integrate the whole PMT signal over $\approx 30$~$\mu$s, to profit of both
the fast and slow
components of LAr. 
For each TPC chamber, the resulting integrated signals were summed
up by a linear analog adder and then discriminated to generate a trigger gate.
The obtained results demonstrate that the implemented trigger system was
effective for a wide range of event energies throughout the CNGS run in a stable way~\cite{Antonello_2014}.

\subsection{The 2.3~L WARP detector prototype}

A direct search of Dark Matter in the form of Weakly Interacting Massive Particles
(WIMP) was carried out at LNGS with a small scale 
LAr-TPC prototype detector to prove the effectiveness of this
detection technique~\cite{BENETTI2008495}. The detector was
based on a small 2.3~L test chamber filled with LAr. The scintillation
light was detected by a set of 7 PMTs with a diameter of 2~inches.

The trigger system was based on a hybrid solution. At first level  
PMT signals coming from the last dynode were discriminated with a threshold of 1.5 photoelectrons.
The {\em Digital Trigger}\/ condition was a coincidence of at least 3 PMTs corresponding 
to about 3.5 keV ion recoil energy. 

\begin{figure}[h]
\centering
\includegraphics[width=0.95\columnwidth]{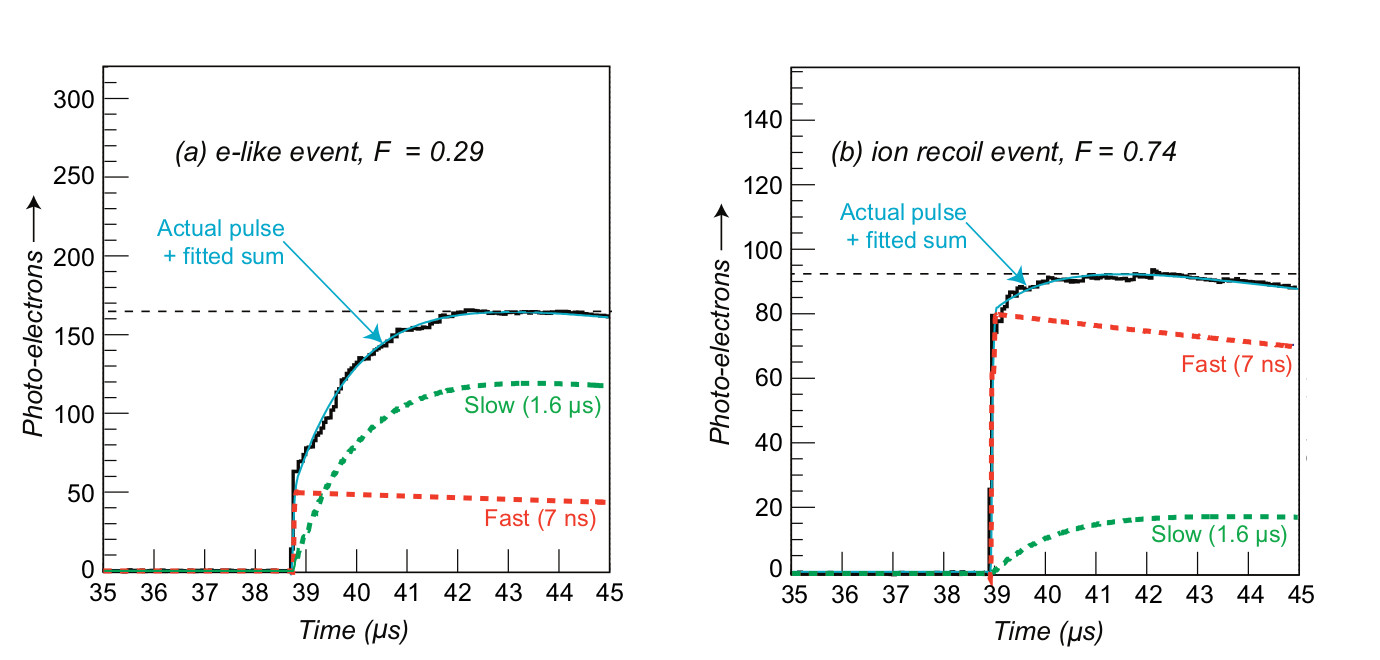}
\caption{2.3~L WARP prototype: typical e-like and ion recoil signals resulting from the integration and sum up of pulses coming from the anode of 7 PMTs.}
\label{fig:PDF}
\end{figure}

The signal from the anode of each PMT
was integrated by preamplifiers with a time constant of 40~$\mu$s to exploit both
components of the scintillation light.
The preamplifier outputs were digitized and summed up to get signals proportional
to the deposited energy, realizing a second level {\em Analog Trigger}\/ able to
distinguish particle interactions by means of the PSD technique. A fast discrimination parameter $F$, computed as 

$$F = 0.99(A_1/A_2) - 0.118$$ 
where $A_1$ and $A_2$ 
are the integrated signal amplitudes
after 200~ns and 5.5~$\mu$s, allowed a pulse shape discrimination between nuclear recoils
($F \approx 0.75$) and background electrons ($F \approx 0.31$),
resulting  in a strong discrimination against the background (see figure~\ref{fig:PDF}).

\section{The analog linear adder}
\label{sec:board}

\begin{figure}[h]
\centering
\includegraphics[width=0.55\columnwidth]{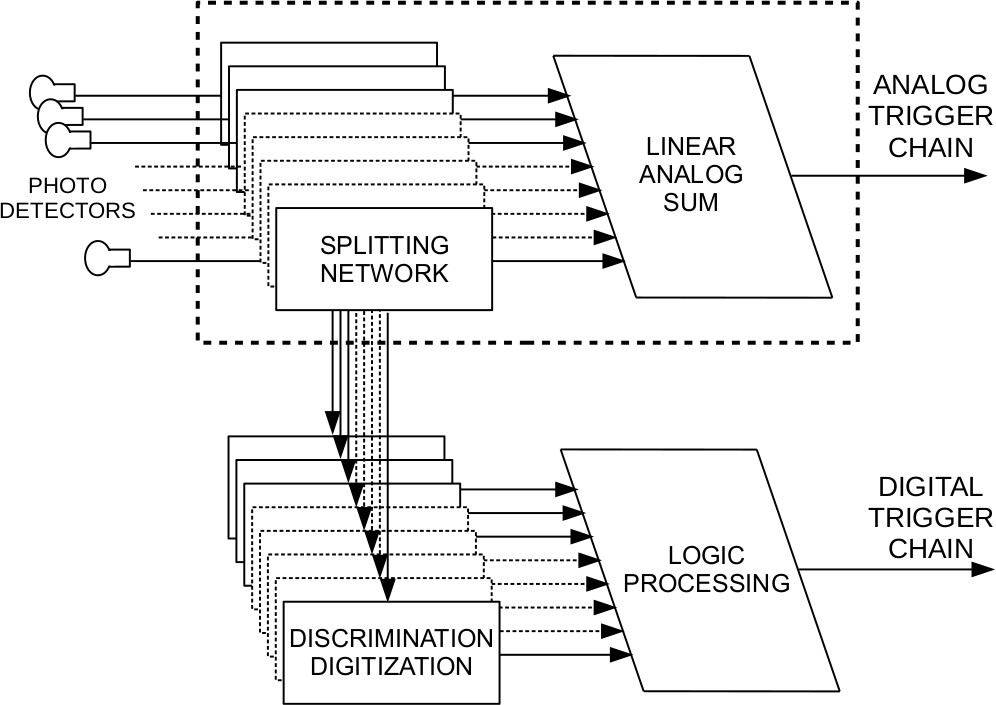}
\caption{Diagram of the adoption of a hybrid trigger solution.}
\label{fig:trigger3}
\end{figure}

The above considerations suggest the adoption at trigger level of hybrid solutions in future large volume LAr-TPC, such as the ICARUS
T600 detector in the SBN program~\cite{Antonello:2015lea}.
For this purpose, an innovative electronic board is being developed in the laboratories of the University of Pavia and INFN, to be used in large volume LAr-TPC 
detectors. The basic idea is to include in a single electronic board a signal splitter and an analog linear adder. 
The output, once properly shaped, digitized and/or discriminated, is used for the generation of an {\em Analog Trigger}\/ gate, while the split signals are used to feed a {\em Digital Trigger}\/  chain as shown in figure~\ref{fig:trigger3}.

\begin{figure}[h]
\centering
\includegraphics[width=0.75\columnwidth]{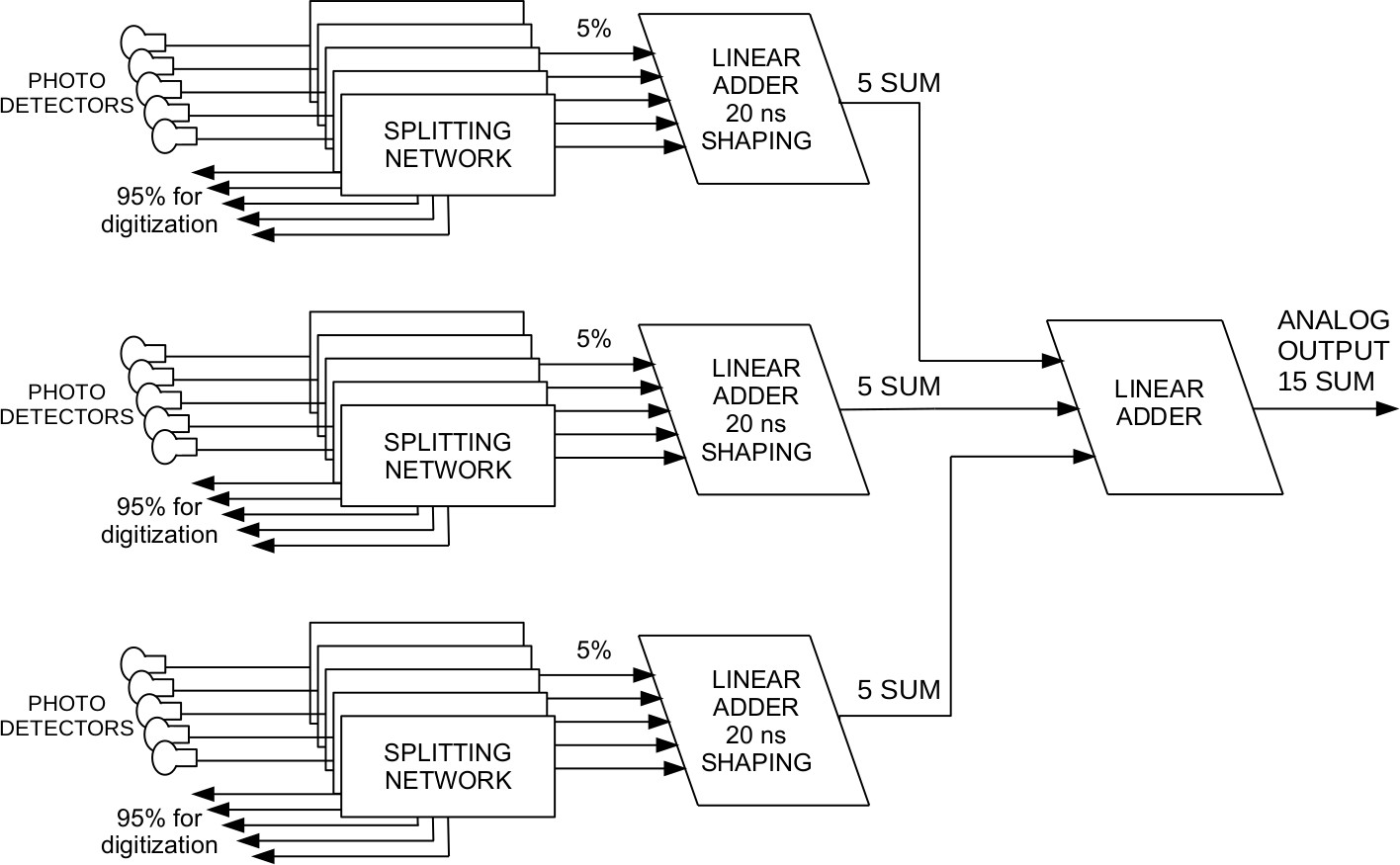}
\caption{Analog linear adder scheme.}
\label{fig:adder}
\end{figure}

\begin{figure}[h]
\centering
\includegraphics[width=0.75\columnwidth]{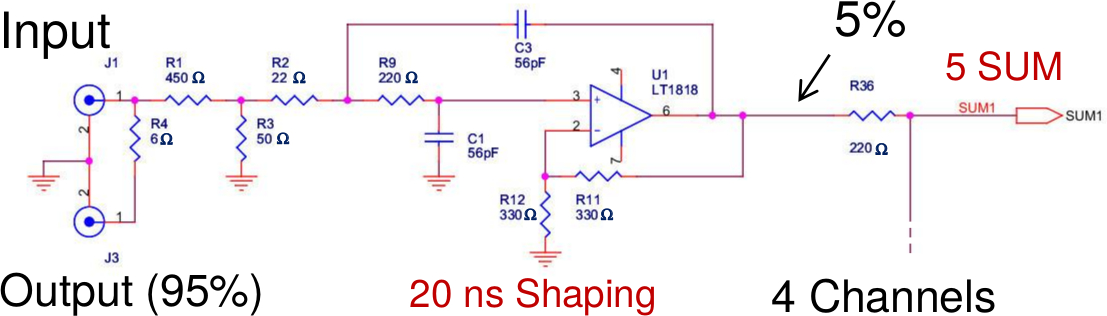}
\caption{Electronic diagram of a 5-channel branch.}
\label{fig:5channels}
\end{figure}

\begin{figure}[h]
\centering
\includegraphics[width=0.75\columnwidth]{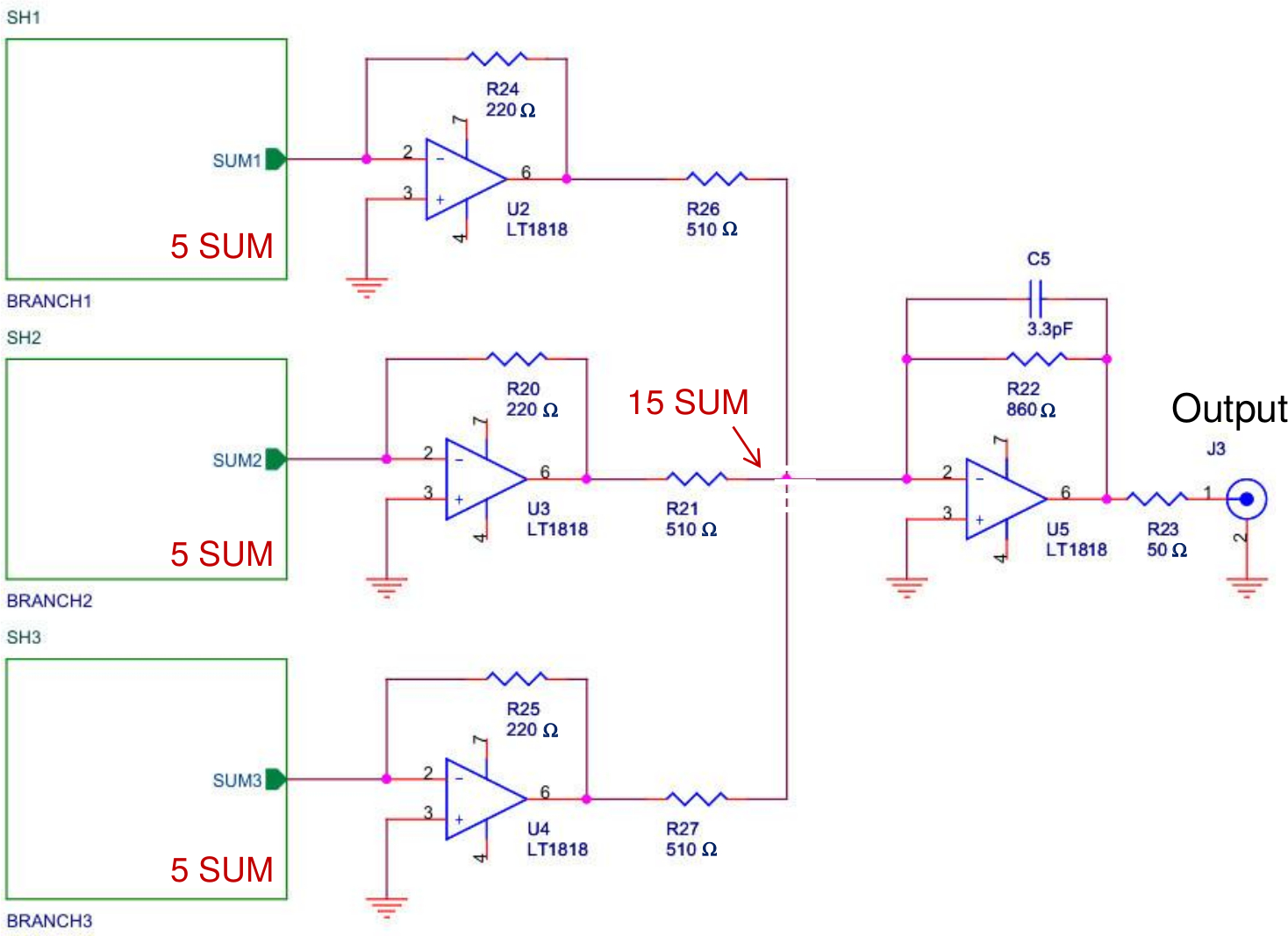}
\caption{Electronic diagram of the complete 15-channel analog adder.}
\label{fig:15channels}
\end{figure}

The analog linear adder consists of 3 independent branches, each summing up the signals coming from 5 independent channels, according to the scheme shown in figure~\ref{fig:adder}.
In each input channel the signal is split by a passive network towards two separate lines which, preserving a 50~$\Omega$ impedance, share
the 5\% and 95\% of the input signal amplitude. In each branch, as shown in figure~\ref{fig:5channels}, each 5\%-line feeds a LT1818 preamplifier used in
Sallen-Key configuration to perform a 20~ns shaping to partially overcome the time spread among different input channels.
Signals coming from the 3 branches are in turn summed up by a 
secondary linear adder realizing
a 15-channel analog sum (figure~\ref{fig:15channels}). 
In parallel, the 15 lines delivering the 95\% of the input signal amplitude are available for fast digitization and signal recording.

The basic component of the analog linear adder is the LT1818 amplifier, employed as buffer to drive 50~$\Omega$ loads and
to perform Sallen-Key shaping of the input signals.
The board power supply voltage is $\pm 5$~V
which permits a good output linearity in the $\pm 2$~V output range.

\begin{figure}[h]
\centering
\includegraphics[width=0.55\columnwidth]{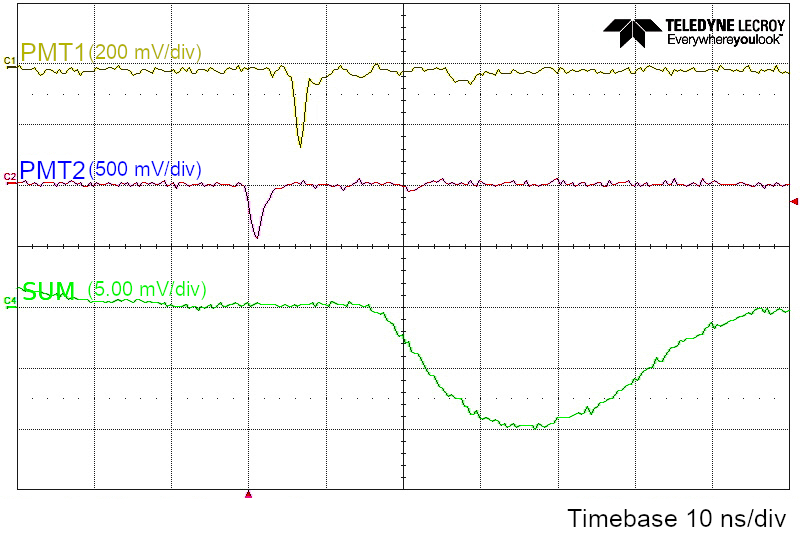}
\caption{Example of resulting analog sum of two actual PMT
channels processed by the electronic board prototype described in the paper.}
\label{fig:lecroy}
\end{figure}

\begin{figure}[h]
\centering
\includegraphics[width=0.45\columnwidth]{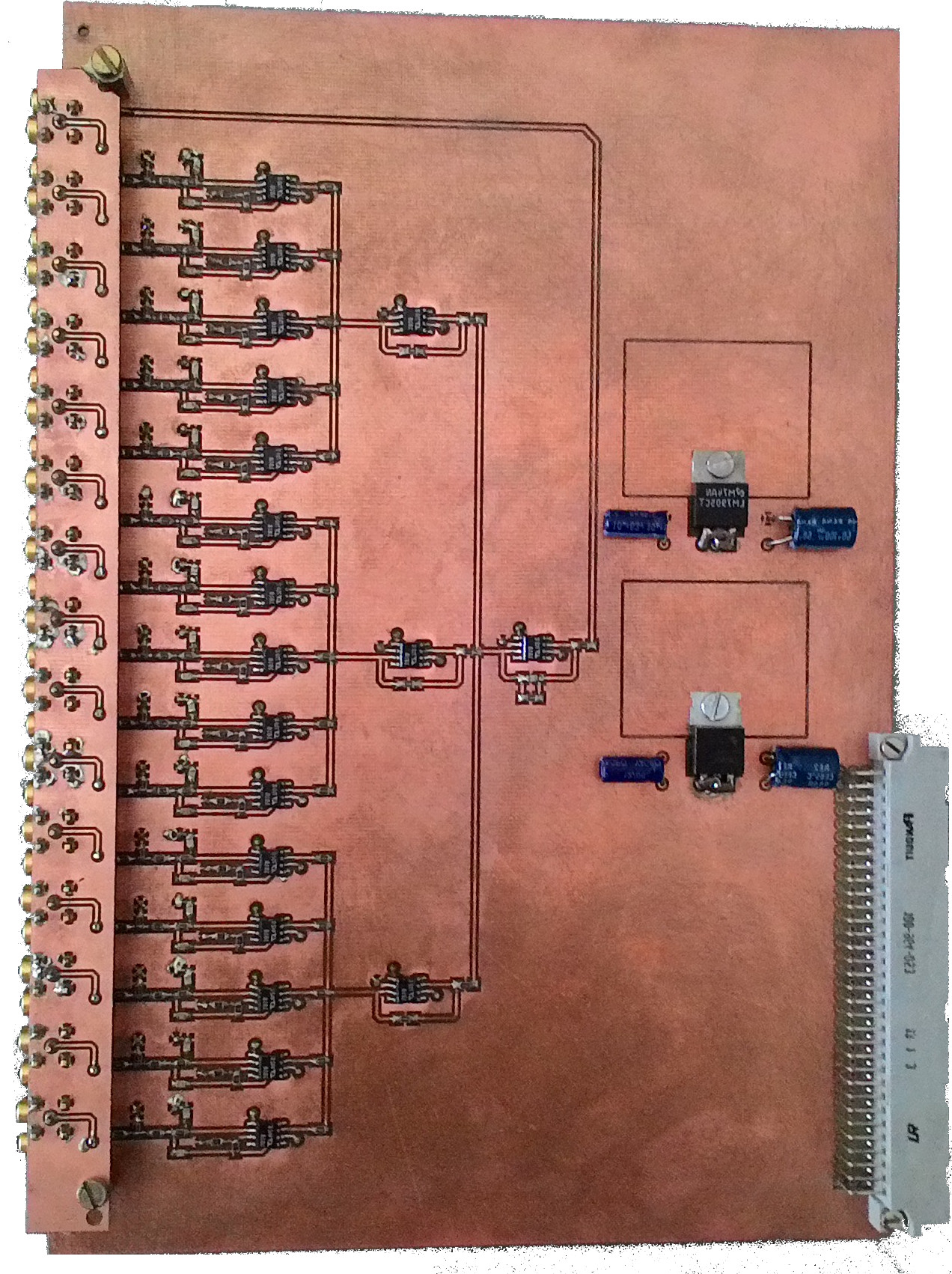}
\caption{Picture of the final analog adder prototype. The different circuit stages described in the text
are easily recognizable in the circuit.}
\label{fig:prototype3}
\end{figure}

The electronic scheme was extensively simulated by means of the LTspice\footnote{LTspice is a freeware computer software implementing a SPICE electronic circuit simulator, produced by semiconductor manufacturer Linear Technology Corp., now part of Analog Devices Inc.} package and a NIM prototype was realized and tested 
at CERN using a 10-PMT test facility~\cite{BABICZ2019162421,RASELLI}. 
The board behavior is shown, as example, in figure~\ref{fig:lecroy}, where the resulting analog sum of two actual PMT channels is presented.  For the final production a standard 6U-VME assembly is adopted, as shown in figure~\ref{fig:prototype3}. Since a large number of connectors (31) finds place on the 6U-4TE front panel, the MCX standard is employed.

\section{Conclusions}

Two different trigger approaches are presently adopted to provide LAr-TPC detectors with an effective trigger system based on the detection
of the LAr scintillation light. The two techniques are based on analog or digital processing of signals coming from photodetectors.

The adoption at trigger level of hybrid solutions is suggested for future large
volume LAr-TPC. For this purpose
an innovative custom-designed analog board is being designed. This board permits the achievement of a linear analog sum of the photodetector signals and in parallel allows the fast digitization of the whole shape.

\section*{Acknowledgment}
The realization of the electronic board presented in this work is funded by INFN in the framework of the 
ICARUS program finalized to the overhauling of T600 detector in view of its operation on SBN at Fermilab.

\bibliography{bibfile}

\end{document}